\documentclass[final,3p,times,twocolumn]{elsarticle} 

\usepackage{amssymb}
\usepackage{amsmath}

\AtBeginDocument{
 \newwrite\bibnotes
 \def\bibnotesext{Notes.bib}
 \immediate\openout\bibnotes=\jobname\bibnotesext
 \immediate\write\bibnotes{@CONTROL{REVTEX41Control}}
 \immediate\write\bibnotes{@CONTROL{
apsrev41Control,author="08",editor="1",pages="1",title="0",year="1"}}
 \if@filesw
 \immediate\write\@auxout{\string\citation{apsrev41Control}}
 \fi
}

\usepackage{amsmath}
\usepackage{mathscinet}
\usepackage{amssymb}
\usepackage{fancyhdr}
\usepackage{emptypage} 
\usepackage[toc, title]{appendix}
\usepackage{tabularx}
\usepackage{graphicx}
\usepackage{multirow, array}
\usepackage[utf8]{inputenc}
\usepackage{indentfirst}
\usepackage{hyperref} 
\usepackage[table]{xcolor} 
\usepackage{ulem} 
\usepackage{color}
\usepackage{url} 

\usepackage{eurosym}
\usepackage{longtable}
\usepackage{makecell}
\usepackage{xr}
\usepackage{chngcntr}
\usepackage{animate}
\newcommand{\tf}{t_{\text{f}}}
\newcommand{\nf}{n_{\text{f}}}

\usepackage{comment}

\usepackage{chngcntr}

\usepackage{caption}

\counterwithin{figure}{section}
\counterwithin{table}{section}

\begin{document}

\begin{frontmatter}

\title{An Analytics Framework for Modeling Residential Photovoltaic Adoption and Decision Dynamics} 

\author[1]{Canig\'o Callau-Boix} 
\ead{canigo@ifisc.uib-csic.es}

\author[1]{Ra\'ul Toral\corref{cor1}}
\ead{raul@ifisc.uib-csic.es}
\cortext[cor1]{Corresponding author}

\author[1]{Pere Colet}
\ead{pere@ifisc.uib-csic.es}

\affiliation{organization={Institut de F\'isica Interdisciplin\`aria i Sistemes Complexos IFISC (CSIC-UIB)},
 addressline={Campus Universitat de les Illes Balears}, 
 city={Palma de Mallorca},
 postcode={07122}, 
 state={Illes Balears},
 country={Spain}}

\begin{abstract}
Photovoltaic generation plays a central role in the energy transition, yet understanding its adoption dynamics requires robust analytical frameworks that capture both temporal and spatial patterns of decision behavior. This study applies a data-driven decision analytics approach to examine residential self-consumption photovoltaic installations in Catalonia within an innovation diffusion framework. The temporal evolution of adoption is modeled using a logistic growth function, providing evidence that imitation effects are a primary driver of adoption decisions. To extend the analysis, a quantitative methodology is developed to estimate the influence of external factors on adoption behavior, revealing that social perception exerts a stronger impact than regulatory and socioeconomic variables when considered independently. In addition, a spatial analytics component is incorporated to assess territorial heterogeneity, identifying correlations between adoption patterns and demographic and socioeconomic characteristics. The findings contribute to predictive and diagnostic analytics by offering a structured framework to model technology diffusion and inform policy and investment decisions aimed at accelerating sustainable energy adoption.
\end{abstract}


\end{frontmatter}


\externaldocument{supplementary_v2}

\section{Introduction} \label{intro}

Driven by the global transition toward renewable energy sources~\cite{ipcc}, photovoltaic (PV) generation has experienced rapid growth and is expected to continue expanding in the coming years. This expansion is largely attributed to significant improvements in conversion efficiency, which have progressively reduced costs and made PV energy increasingly competitive with fossil fuels. In addition, PV generation can be efficiently implemented on a wide range of scales, from large solar farms spanning several kilometers to small residential installations consisting of only a few panels.

In this paper, we focus on the diffusion of residential photovoltaic installations for self-consumption, owing to their distinctive social and economic implications. By enabling households to generate their own electricity, residential PV systems transform consumers into “prosumers”, reducing their dependence on centralized utilities, cutting electricity costs, and gaining long-term financial predictability. Empirical evidence supports these benefits. For instance, analyses of self-consumption systems in Spain show that PV adoption can yield substantial lifetime savings~\cite{Garcia-Lopez2023, Bertsch2021, Menezes2017}. Similarly, research on adopters in the United States indicates that electricity cost savings depend strongly on local policies and system characteristics, highlighting the role of regulatory frameworks and market design in shaping the private economic returns of residential PV investments~\cite{Fikru2020}.

Beyond economical implications, widespread residential PV helps democratize the energy system itself. Instead of relying solely on large, centralized power plants owned by a few corporations, energy generation becomes distributed across millions of homes and communities. Residential PV is, thus, integral to discussions about energy access, justice, and household engagement with the energy transition. 

A systematic review of household-level solar adoption~\cite{Shakeel2023} identifies a broad set of factors influencing this process, which can be grouped into four main categories: economic factors (e.g., costs), environmental awareness, market conditions, and the regulatory and legislative framework.

Besides these factors, interactions between individuals are known to play a key role. A survey performed in Texas (USA)~\cite{rai2013effective} found that peer effects, either passive through witnessing PV systems in the neighborhood or active through peer-to-peer communication, significantly decreased decision times. 
The authors of this study also found that contact with neighbors before installation was the single most effective strategy for speeding decision times. Along the same line, surveys performed in Sweden point to the importance of direct contacts~\cite{palm2017peer, mundaca2020drives}, primarily to confirm that the technology works and its acquisition is viable. 

In this context we aim to analyze the mechanisms behind the decision of individuals to adopt residential PV, with particular attention to the role of social interactions, as well as the influence of the broader socio-economic and fiscal environment. 

To this end, we apply the diffusion of innovation framework which seeks to explain how new ideas and technologies spread in human societies. Starting in late 19th century as a branch of sociology, the theory was developed and popularized by Everett Rogers in his book \textit{Diffusion of Innovations}~\cite{rogers1962}, published in 1962. From a quantitative perspective, diffusion of innovations has been described first in terms of the logistic model~\cite{mansfield1961technical}, and later by a model introduced by Bass in 1969~\cite{Bass:1969}. The Bass model was subsequently generalized to include external influences like advertising or government regulations~\cite{Bass1994}. The Bass model has been applied as analytical framework for innovation across a wide range of contexts ranging from medical technology \cite{Peters2025} to military technology and combat strategy \cite{Schuur2025}. The simpler logistic model is still very useful in scenarios where imitation plays a major role, including a recently introduced predictive modeling framework for sales of electric and autonomous vehicles \cite{Alatawneh2024}.

In the photovoltaic sector, the Bass model and its extensions have been widely applied to describe and forecast market evolution across different countries~\cite{guidolin2010cross, agarwal2015model, wang2017model}, as well as to assess the impact of government incentives on adoption dynamics~\cite{Bunea2020}. However, most of these studies do not distinguish between large-scale commercial solar plants and residential self-consumption installations, despite the fact that the underlying decision processes and driving factors may differ substantially. Although Ref.~\cite{agarwal2015model} specifically examines residential installations, it relies on a dataset covering only a single year, which limits the possibility of empirically identifying the effects of tax and regulatory changes over time.

Our study focuses on the Spanish autonomous community of Catalonia and spans the full evolution of photovoltaic adoption, from its early stages in 2002 to 2024. The dataset comprises hundreds of thousands of geolocated installations, enabling the identification of distinct growth phases. Our analysis shows that, for residential self-consumption PV systems, the innovation parameter in the Bass model plays a negligible role; instead, growth is primarily driven by imitation and is well described by a logistic model. The estimated parameters further indicate that the inflection point has already been reached and that installations are approaching saturation.

Beyond the fitting of the models, we propose an original methodological approach that enables the empirical estimation of external influences without imposing a priori assumptions on their functional form. In particular we analyze the effect of external regulatory and socio-economical factors and found that, more relevant than these factors by themselves, is the social perception of them. 
In addition, we conduct a spatial analysis of adoption dynamics, examining their correlation with a range of socioeconomic and demographic variables to better understand the territorial heterogeneity of residential PV diffusion.

This paper is structured as follows: Section~\ref{Data} provides an overall discussion of the data sets used in this work. Section~\ref{Size_distribution} discusses the characteristic size of installations. Section~\ref{Modelization} discusses logistic and Bass models and their fit to the data. Section~\ref{External_factors} discusses the influence of external factors on the evolution. Section~\ref{early_evolution} discusses the early evolution with a very small number of installations. Section~\ref{Spatial_analysis} provides an spatial analysis of installations and its relation with different socioeconomic indicators. Section~\ref{Conclusions} provides concluding remarks.

\section{Data}
\label{Data}

The main dataset used in this study is the \textit{Registre d'Autoconsum a Catalunya} (RAC), available upon request at the \textit{Generalitat} web page~\cite{sollicitud_rac}. It includes all generation installations connected to the grid whose main purpose is self-consumption (also known as \textit{behind-the-meter} installations) while installations intended for wholesale/utility supply (\textit{front-the-meter} installations) are not included~\cite{RD2019} (see the supplementary material \ref{supplementary:dataset_info} for more details). A version of this register without coordinates is publicly available~\cite{rac_public}. Up to May 31, 2024 there were a total of $112,526$ installations. After excluding non-photovoltaic systems (such as wind turbines or biogas) and repeated entries at the same location we ended up with a dataset including $112,086$ installations.

We also made use of other datasets and information to analyze the dependence of installations data with different types of variables:
\begin{itemize}
 \item Demographic: population by municipalities~\cite{pobmun} and by \textit{comarques} (aggregations of several municipalities) in 2023~\cite{pobcom}, population density by municipalities considering only urban soil in 2024~\cite{denmun}, 
 \item Urban planning: statistics of the number of stories in buildings~\cite{n_plantes}, and spatial distribution of buildings in Catalonia in 2013~\cite{corral2022finite,corral_private_comm}.
 \item Economic: electricity price in Spain~\cite{pvpc}, gross disposable household income (GDHI) by municipalities~\cite{rfdbmun}.
 \item Regulatory: subsidies to renewable sources installations~\cite{subvencions} from \textit{Generalitat de Catalunya} (Catalan government) and of tax deductions on the personal income tax (Spanish IRPF)~\cite{irpf}.
 \item Social concern: searches in Google of "impuesto al sol" (\textit{Solar tax})\cite{trends_impost} and of "precio luz" (\textit{price of electricity})\cite{trends_preu_llum} in Spain.
 \item Geographical: Maps of Catalonia by different scales of aggregation~\cite{mapes_cat}, of solar irradiation in Catalonia~\cite{atlas_solar}, and of states of the world~\cite{mapes_mon}.
\end{itemize}

\section{Distribution of capacity per site: Residential vs non-residential}
\label{Size_distribution}

\begin{figure}
\includegraphics[width=\columnwidth]{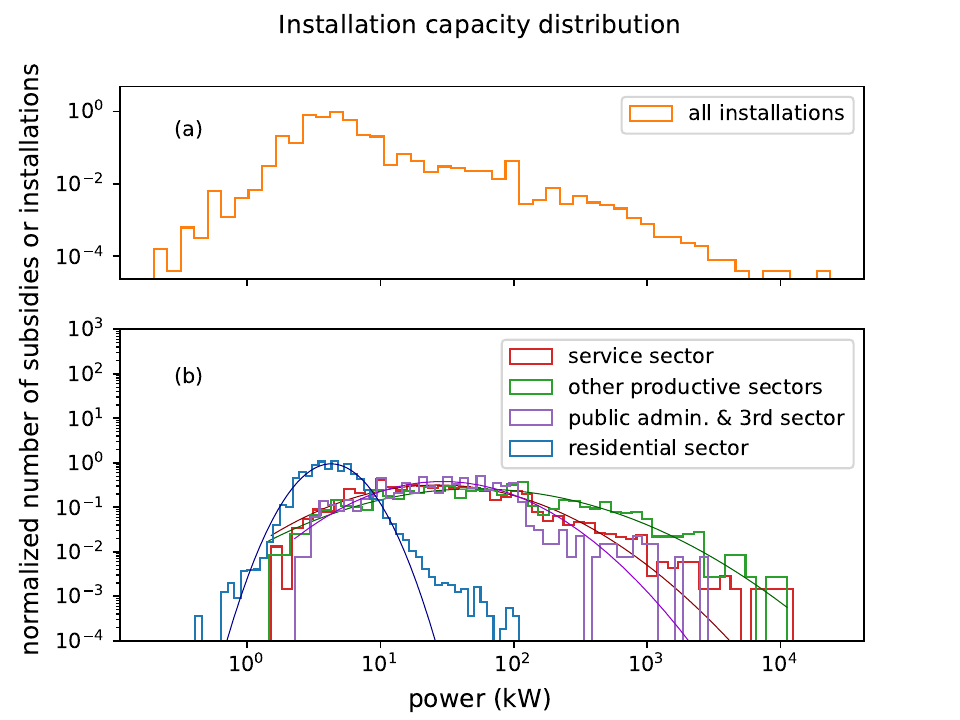}
\caption{(a) Distribution of the capacity per site with log-spaced bins with a double logarithmic scale from the RAC dataset. (b) Distribution of the capacity per site for the subsidy programs shown in Table~\ref{table:sectors} and the corresponding lognormal fits, also in double logarithmic scale.}
\label{figure1}
\end{figure}

The distribution of the installed capacity per site obtained from the RAC dataset, using logarithmic binning, is shown in Fig.~\ref{figure1}a. It spans from small residential installations of $0.2$~kW up to large commercial ones of $11,820$~kW. 
The distribution is clearly asymmetric with a longer tail on the right, a main peak around $5$~kW, and a smaller secondary peak at $100$~kW followed by a drop. 
The bimodal structure suggests that that residential installations exhibit a characteristic scale, around the location of the main peak while the secondary peak is associated to non-residential installations. The drop after the secondary peak can be understood from the fact that, according to Spanish legislation~\cite{RD2019}, surplus compensation (being financially credited on the electricity bill for the excess generation fed into the grid) is limited to installations smaller than $100$~kW.

The RAC dataset does not contain information about the sector to which an installation belongs. Therefore, to analyze the typical capacity in different sectors we resource to the \textit{Generalitat} subsidies dataset~\cite{subvencions} which is smaller but has the advantage of being disaggregated in programs that cover different sectors, as listed in Table~\ref{table:sectors}). Within program $4$ we further disaggregate the subsidies granted to individuals (residential installations) from those granted to public administrations and third sector.
The distribution of capacity per site for each sector can be adjusted to a log-normal distribution as shown in Fig.~\ref{figure1}b with parameter estimates listed in Table~\ref{table:sectors}.
The distribution for the residential sector has much smaller mean ($4.7$~kW) and standard deviation ($2.7$~kW) than for other sectors, corroborating the existence of a characteristic scale. In particular $97.6\%$ of the residential installations have a capacity below $10$~kW and $97.2\%$ of the subsidized installations smaller than $10$~kW are residential. 

\begin{table*}
 \centering
 \begin{tabular}{ | {c} | m{2.8cm} | r r | r r | r r | }
 \hline
 prog. & \multicolumn{1}{c |}{sector} & \multicolumn{2}{c |}{subsidized install.} & \multicolumn{2}{c |} {capacity (kW)} & \multicolumn{2} {c|} {lognomal} \\
 & &sector & $<10$kW & mean & st. dev. & \multicolumn{1}{c }{$\mu$} & \multicolumn{1}{c |}{$\sigma$} \\
 \hline
 1 & service & 3795 & 919 & 79.6 & 365.0 & 3.291 & 1.257 \\
 2 & other & 2026 & 299 & 184.7 & 520.2 & 3.942 & 1.538 \\
 4 & public adm. \& third & 922 & 153 & 57.1 & 147.1 & 3.372 & 1.048 \\
 4 & residential & 49022 & 47846 & 4.7 & 2.7 & 1.452 & 0.421 \\
 \hline
 \end{tabular}
\caption{Statistical data for \textit{Generalitat} subsidies addressed to new PV installations. Lognormal parameters are computed using a logarithmic scale in base e for the installation power measured in kW.}
\label{table:sectors}
\end{table*}

From now on, we focus on the RAC dataset. The distribution observed in the subsidies dataset supports the interpretation that the double peak identified in the RAC dataset (Fig.~\ref{figure1}a) arises from classifying installations with a capacity below $10$~kW as residential, while larger systems are attributed to other sectors. Using this threshold value yields $102,858$ ($91.77\%$) residential installations and $9,228$ ($8.23\%$) in other sectors. However, it is interesting to note that, compared with the total installed capacity ($1,172$~MW as of May 31, 2024), non-residential installations account for $59.8\%$ of it ($701$~MW), way over the residential contribution of $40.2\%$ ($471$~MW). 

\begin{figure}
\includegraphics[width=\columnwidth]{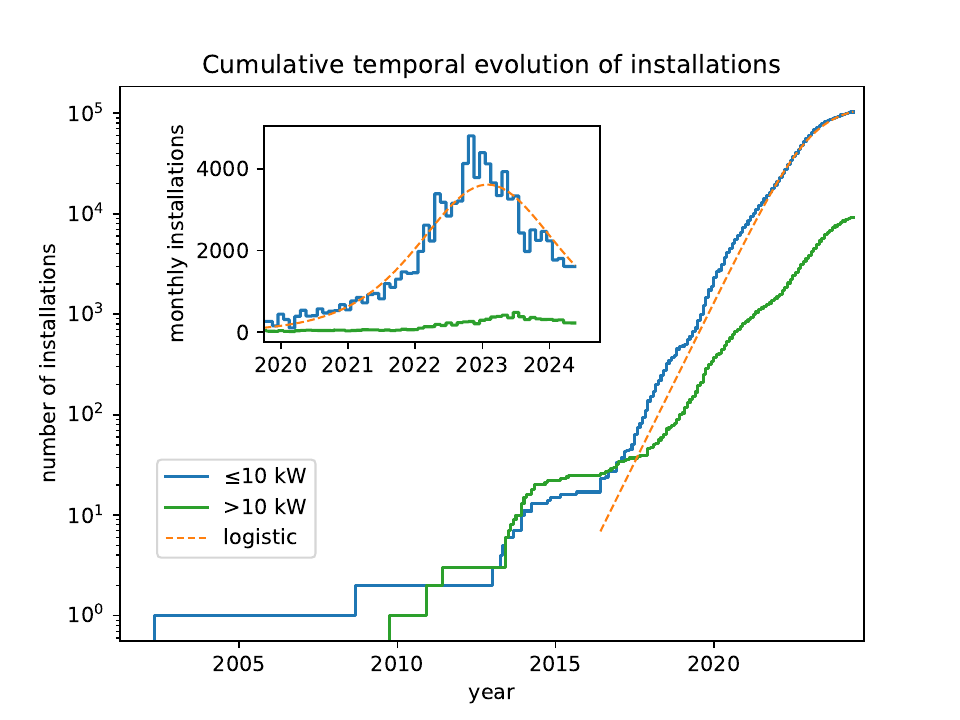}
\caption{Time evolution of the cumulative number of installations in logarithmic scale disaggregated in residential (blue) and non-residential (green). The inset shows the monthly installations for each category (not stacked). The dashed orange line shows the fit using the logistic model from June 2016 to May 2024 as discussed in Section~\ref{Modelization}.}
\label{fig_install_evolution}
\end{figure}

The main panel of Fig.~\ref{fig_install_evolution}, shows the evolution of the cumulative number of residential (blue) and non-residential (green) installations. The early evolution, up to June 2016, is discussed in section~\ref{early_evolution}. Afterwards, starting June 2016 until 2022 the growth is approximately exponential for both residential and non-residential installations. In 2023 there is a slowdown signaling a saturation effect. This trend is also evident in the monthly number of new installations shown in the inset, which reaches a maximum value. For residential installations the maximum takes place earlier and it is sharper than for non-residential ones.

\section{Modelization of the temporal evolution} \label{Modelization} 

We model the evolution of the number of PV installations as a set of $N$ agents which can adopt a new technology. $N$ plays the role of a market capacity and is assumed to be constant in time. We also assume that agents that have installed do not uninstall, so that the number of installations $n(t)$ is a monotonically increasing function of time. Its evolution is given by:
\begin{equation} \label{eq:n(t)}
 \frac{dn(t)}{dt}=\omega(n(t))(N-n(t)),
\end{equation}
where $\omega(n)\ge 0$ is the adoption rate.

Starting from $n(t_0)<N$, as $n(t)$ grows and approaches $N$, its growth rate $dn/dt$ approaches to $0$ (saturation effect). For large times the system reaches a stable fixed point $n=N$ in which all agents have adopted the technology. 
The dependence of the adoption rate $\omega(n)$ on the state of the system defines the different models:
\begin{itemize}
 \item The logistic model (LM), first introduced by Verhulst~\cite{Verhulst1838, Verhulst1845} for population growth, has been widely used in different fields, including diffusion of innovations~\cite{mansfield1961technical}. It considers contagion-like interactions in which non-adopter agents can imitate adopters. The adoption rate is proportional to the fraction of adopters, namely 
 \begin{equation}
 \omega(n)=qn/N,
 \label{eq:omega_logistic}
 \end{equation}
 where $q$ is a parameter of the model.
 Introducing Eq.~(\ref{eq:omega_logistic}) in Eq.~(\ref{eq:n(t)}) and integrating yields
 \begin{equation}
 n(t)=\frac{N}{1+\exp(-q(t-t_{\rm I}))},
 \label{eq:S_logistic}
 \end{equation}
 where $t_{\rm I}$ is the time at which $n(t)$ displays an inflection point (maximum of the growth rate $dn/dt$), at which $n(t_{\rm I})=N/2$. Note that since $n=0$ is an unstable fixed point of Eq.~\eqref{eq:n(t)} this model needs a non-zero initial value to start growing.

 For $t \ll t_{\rm I}$ the growth of $n(t)$ can be approximated by an exponential function:
 \begin{equation}
n(t) \approx \exp(q(t-t_1)).
 \label{eq:S_exponential}
 \end{equation}
 where $t_1$ is the time of the first installation. 
 
 \item The Bass model (BM)~\cite{Bass:1969} is an extension of the logistic, which adds a constant term $p$ to the adoption rate:
 \begin{equation}
 \omega(n)=p+qn/N.
 \label{eq:omega_Bass}
 \end{equation}
 The model interprets $p$ as an innovation coefficient, which represents the will of agents to adopt the technology due to external influences independently of what other agents do (advertising, mass media, imitation from outside the set under study, etc.)~\cite{BERTOTTI201655}. Meanwhile, $q$ is interpreted as an imitation coefficient, corresponding to contagion-like social interactions as in the LM. Integrating Eq.~(\ref{eq:n(t)}) with the rate gieven by Eq.~(\ref{eq:omega_Bass}) yields
 \begin{equation}
 n(t)=N\frac{1-\frac{p}{q}\exp(-(p+q)(t-t_{\rm I}))}{1+\exp(-(p+q)(t-t_{\rm I}))}
 \label{eq:S_Bass}
 \end{equation}
 where $t_{\rm I}$ is the time at which $n$ has an inflection point in which the growth rate has a maximum and $n(t_{\rm I})=\frac{N}{2}(1-\frac{p}{q})$. Thanks to innovation term, growth can start from $n=0$.
\end{itemize}

In general, diffusion processes can be influenced by a wide range of factors~\cite{wejnert2002integrating}, many of which may lead to similar mathematical behaviors. However, as long as the adoption rate can be approximated to an affine function of the number of adopters, the Bass model can still be applied using effective parameter values.

We now fit the model parameters using the cumulative number of residential installations from June 2016 onward. \footnote{Parameters has been fitted using the \textit{curve\_fit} function of scipy.optimize Python library using the trust region reflective algorithm and setting bounds to limit estimates to positive values~\cite{curve_fit}. This function returns the optimal set of parameters along with their associated covariance matrix, from which the standard deviation errors of the parameters can be computed.}

The results for the logistic model are shown as a dashed orange line Fig.~\ref{fig_install_evolution}, with the corresponding parameter estimates displayed in Table~\ref{table:fits}. The fitting is very good, despite some deviations in the early part of the considered period with less than $1000$ accumulated installations. Fitting of the Bass model yields a value for $p$ consistent with zero, implying that innovation turns out to be irrelevant and therefore Bass and logistic models describe equally well the growth from June 2016. In this situation simpler models are preferred since they facilitate their interpretation~\cite{ormerod2006validation}.
On a quantitative basis, it is possible to use the Bayesian Information Criterion (BIC)~\cite{neath2012bayesian} for model selection. It is defined as $\text{BIC}=k\ln(n)-2\ln(L)$, where $k$ is the number of parameters estimated by the model, $n$ the number of data points, and $L$ the maximized value of the likelihood function of the model. The lower $\text{BIC}$ is, the better, and differences larger than 2 are considered significant. The obtained values are $\text{BIC}_\text{LM}=816.9$ nats and $\text{BIC}_\text{BM}=821.5$ nats, what indicates the logistic is significantly more adequate, as also observed for other countries~\cite{guidolin2010cross}.

According to the fitting (Table~\ref{table:fits}) the inflection point occurred at the end of January 2023, and the saturation capacity is $N=118,600$. This capacity is only $6\%$ larger than the $112,086$ installations up to May 31, 2024, therefore the growth is quickly approaching saturation. We would like to remark that good parameter estimation is only feasible if there is data available after the maximum on the growth rate has been reached~\cite{van1997bias}, which is the case here (see the supplementary material \ref{supplementary:robustness} for details about the robustness of the fits).

\begin{table*}
 \centering
 \begin{tabular}{ m{2cm} m{2cm} m{2cm} m{2cm} m{3cm} m{2cm}}
 \hline
 model & $N$ (thousands) & $q$ (1/year) & $p$ (1/year) & $t_{\rm I}$ or $t_1$ (year) & \text{BIC} (nats) \\
 \hline
 logistic & $118.6\ (1.2)$ & $1.46\ (2)$ & & $t_{\rm I}=2023.078\ (19)$ & $816.9$ \\
 Bass & $118.6\ (1.3)$ & $1.46\ (2)$ & $0.0000\ (19)$ & $t_{\rm I}=2023.078\ (19)$ & $821.5$\\
 exponential & & $0.955\ (11)$ & & $t_1=2014.58\ (10)$ &\\
 \hline
 \end{tabular}
 \caption{Parameter estimates and their errors for BM and LM obtained from the cumulative number of residential installations from June 2016 to May 2024 and for exponential growth from June 2016 to June 2023. The parameters are, in order, the saturation number of installations, the imitation rate, the innovation rate, and the inflection time except for the exponential which is the fitted time for the first installation. In the last column, the Bayesian information criterion, in nats.}
 \label{table:fits}
\end{table*}

We also considered alternative mechanisms to contagion that could account for the observed dynamics. Previous studies have shown that sigmoidal adoption patterns may also arise from heterogeneity in individuals’ propensity to adopt~\cite{van2004social}, social influence~\cite{young2009innovation}, or social learning processes~\cite{young2009innovation}.

Following the methodology proposed in~\cite{young2009innovation}, we applied several empirical tests to assess the presence of these mechanisms in our data (see the supplementary material \ref{supplementary:mechanisms}). Our results do not provide clear evidence that they play a distinctive role in the case studied here.

\section{Influence of external factors}
\label{External_factors}

To account for the influence of external factors, such as price, advertising or government regulations, Bass and coworkers proposed the generalized Bass model~\cite{Bass1994} which introduces a time dependent function $\chi(t)$ (named by the authors as ``current marketing effort'') multiplying the Bass adoption rate (\ref{eq:omega_Bass}). Owing to the monotonicity of the evolution, $\chi(t)$ must be non-negative, only being zero when there is no increase of adoptions. This generalization can, in fact, be applied to the any model described by Eq.~(\ref{eq:n(t)}) as:
\begin{equation}
 \frac{dn(t)}{dt}=\chi(t)\omega(n(t))(N-n(t)).
 \label{eq:generalized}
\end{equation}
The introduction of $\chi(t)$ preserves the interpretation of the parameters, the overall behavior and the saturation level, only fastening or slowing the evolution as a dynamic rescaling of time. It also guarantees the existence of an analytical solution provided the original model has one and $\omega(n(t))$ has no explicit time dependency.

The dimensionless function $\chi(t)$ can be constructed to model market variables such as price and advertising~\cite{Bass1994} or to model political and technological changes (see for instance~\cite{Guseo2007}). In the context of photovoltaic installations, the generalized Bass model has been used to address cross-country diffusion~\cite{guidolin2010cross}, and effects of government's incentives~\cite{Bunea2020} (see also~\cite{Guidolin2023} for a review on other variants of Bass model).

Rather than the usual aprioristic determination of $\chi(t)$ to model external influences, here we consider an alternative approach in which we do not assume any previous shape for $\chi(t)$. Instead, we infer it from the actual data and discuss to which extend the shape of this empirical $\chi(t)$ can be understood from external influences.

To do so in Eq.~(\ref{eq:generalized}) we consider discrete time steps, $t_n=t_0+n\delta t$ where $t_0$ is the initial time and $\delta t$ is a chosen time step. We discretize symmetrically $\left.dn(t)/dt\right|_{t=t_n}$ as $(n(t_{n+1})-n(t_{n-1}))/2 \delta t$. Isolating $\chi(t_n)$ results in
\begin{equation}
 \chi(t_n)=\frac{n(t_{n+1})-n(t_{n-1})} {2 \delta t \omega (n(t_n)) (N-n(t_n))} .\label{eq:chi_empirical}
\end{equation}
We then fill the empirical data set $\{n(t_n)\}$ in Eq.~(\ref{eq:chi_empirical}).
The resulting $\{\chi(t_n)\}$ can be interpreted as an empirical measure of temporal deviations between the data and the model with parameters given by the overall fit. For a model that fits perfectly the data, $\chi(t_n)=1$ for all $n$. 

\begin{figure}
\includegraphics[width=\columnwidth]{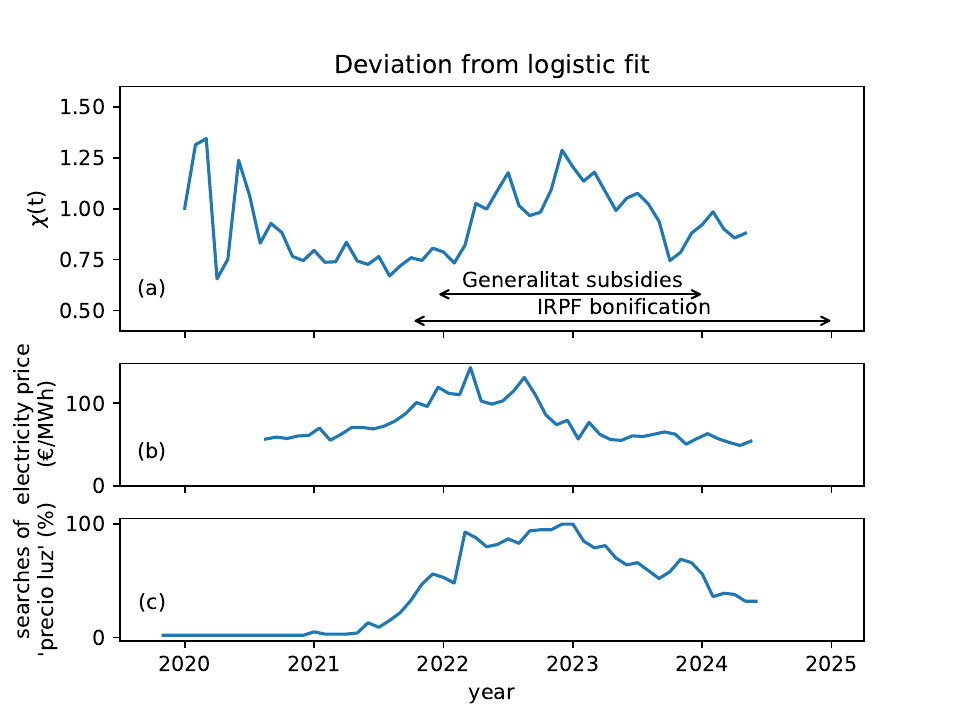}
\caption{(a) Empirical value $\chi(t)$ (November 2019 - May 2024). Double arrows show the periods in which subsidies and personal income tax (IRPF) deductions were in effect. (b) Electricity price in the regulated market according to PVPC (\textit{precio voluntario para el pequeño consumidor}) (August 2020 - May 2024). (c) Normalized number of Google searches of \textit{precio luz} in Spain (November 2019 - May 2024).}
\label{figure4}
\end{figure}

Fig~\ref{figure4}a shows the result for $\{\chi(t_n)\}$ considering the logistic model ($\omega(n)$ given by Eq.~(\ref{eq:omega_logistic})). After an initial regime characterized by a large variability associated to statistical fluctuations due to a relatively small number of monthly installations, $\chi$ takes a value below $1$ during $2021$ which implies a slow evolution. In February $2022$, $\chi$ starts to increase, and in the last part of $2022$ and the first half of $2023$ is larger than $1$, implying a rate of growth faster than the one implied by the original Bass model. This is followed by an overall decrease up to the end of the data. We now proceed to discuss these results in the context of several external influences. 

First, the Catalan government issued a call for subsidies to PV installations on December 14, 2021, which was open until December 31, 2023~\cite{subvencions}. The increase in $\chi$ observed on 2022 agrees quite well with the start of the call: $\chi$ remains above $1$ for most of the time in which the call was active. The decrease in $\chi$ towards the end of the call can be understood from i) the fact that applications overcome the budget well before the end date and ii) the slowness of the adjudication procedure~\cite{subvencions_lentes}. The combination of these two elements has probably discouraged potential applicants by the end the call. Altogether we can conclude that the values of $\chi$ agree qualitatively with the subsidies period. 

Second, the Spanish government has introduced tax deductions on the IRPF for taxpayers improving household energy efficiency, including installation of PV panels, from October 6, 2021 to December 31, 2027~\cite{irpf}. The fact that $\chi$ has an overall decline since mid 2023, while IRPF deductions are in effect, indicates that the effect in the number of adoptions is less relevant than that of direct subsidies. 

Third, the price of the electricity increased on the second half of 2021 and remained quite high until autumn 2023 as shown in Fig.~\ref{figure4}b. This was due to several effects, including the reduced natural gas stock in Europe as consequence of the pandemic on 2020, the increase in CO$_2$ emissions costs on the European Union Emission Trading System (EU ETS), and the fact that the electricity market is marginalist so that the price is determined by the last, most expensive, source that enters in the mix. 

In 2022, following the Russian invasion of Ukraine, uncertainty about supply and sanctions against Russia further drove up gas prices in Europe, directly impacting electricity. In Spain this was exacerbated by the lowest hydroelectric generation in 30 years due to drought~\cite{hydro_Spain}. The price of electricity would have risen even higher were it not for the Iberian exception mechanism, which capped the price for gas used in power plants~\cite{HIDALGOPEREZ2024114092}. This mechanism was applied from June 2022 to February 2023, when gas prices returned to normal. 

As shown in Fig.~\ref{figure4}, the increase in the electricity price precedes that of $\chi(t)$ in about half a year (associated to the time to take the decision to install and the installation time). Furthermore, there is about one year delay between the decrease in the electricity price and that of $\chi(t)$ (associated to the persistence of memories of large prices and installations that were in process). Kendall correlation for the overall period from August 2021 to May 2024 takes a maximum value of $0.69$ at a delay of 10 months. 

More relevant than the price itself is its social perception. We consider the volume of searches in Google of "precio luz" (\textit{electricity price}, in Spanish) as an indicator of the concern for the price of electricity in Spanish society. Results from Google Trends~\cite{trends_preu_llum} are shown in Fig.~\ref{figure4}c. There is a clear correlation with the time evolution of $\chi$. For the period August 2021 to May 2024 the Kendall correlation takes a maximum value of $0.8$ at a delay of 2 months, much stronger and immediate than that of the price.

Some municipalities also offer deductions on real state taxes and on construction permits~\cite{incentius_fiscals} for households installing PV panels. However their amounts are significantly smaller than that of IRPF or direct subsidies, thus they are not discussed here. 

Finally, we would like to note that subsidies and tax deductions reduce installations net cost, and, besides fastening the evolution, can also increase the market capacity~\cite{mahajan1978innovation} (a lower net investment leads to more agents being able to install). An increase in market capacity would be noticeable if more than one peak in the installation rate is observed, corresponding to a saturation followed by a new growth stage, up to a saturation at a higher value and so on. This is not observed in the period from 2021 to 2024, thus there is no clear indication of a market capacity increment.

\section{Early evolution} 
\label{early_evolution}

\begin{figure}
\includegraphics[width=\columnwidth]{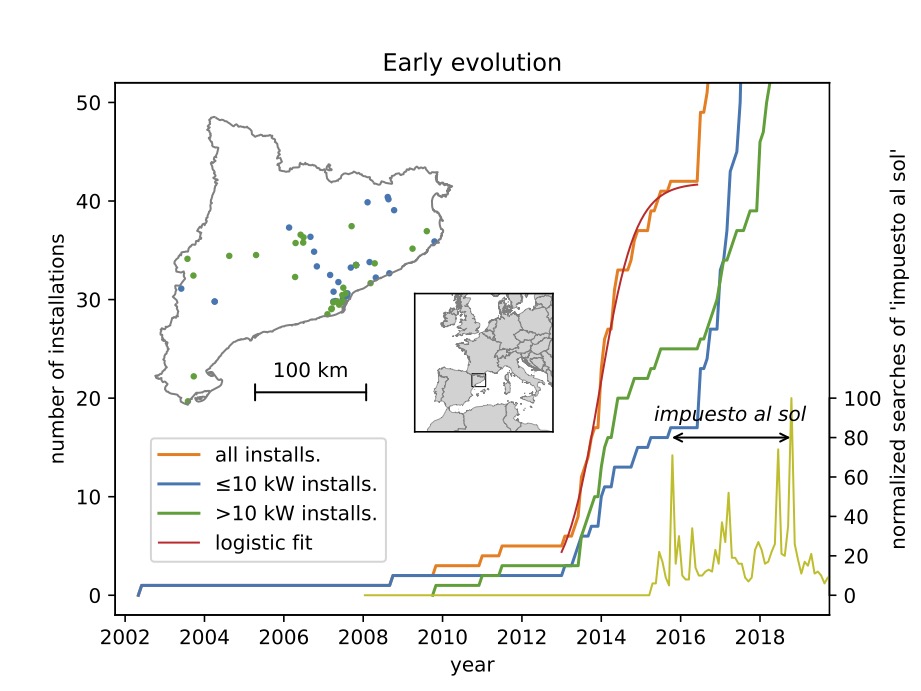}
\caption{Early evolution of PV installations smaller (blue) and larger (green) than $10$~kW and total (orange) as well as the logistic fit from the 5th until the 42th installation. Number of searches of "impuesto al sol" in Spain according to Google Trends relative to their maximum as a function of time (yellow, right axis). The double arrow shows the period in which the tax was in force. The map shows the location of the first 42 installations, separated by color between low and high-power ones.}
\label{fig_early_evolution}
\end{figure}

The first self-consumption PV installation included in the RAC register was installed in May 2002. The initial growth was extremely slow and by December 2012 there were only $2$ installations smaller than $10$~kW and $3$ larger. As shown in Figure~\ref{fig_early_evolution}, starting on 2013 there is a growth period which by September 2015 saturated to a total of $41$ installations, $17$ below $10$~kW and $24$ above. There were no new installations until June 2016 when the growth period discussed in the previous sections started. 
The profile of the cumulative installations is similar to that of logistic growth for both installations smaller (in orange) and larger (in green) than $10$~kW as well as to the total number of installations (in blue).

To analyze this we fit the data for the total cumulative installations from the 5th to the 42th installations in the logistic and Bass models. For the logistic model (in red) we estimate $N=41.8\pm0.4$, $q=2.22\pm0.08\ year^{-1}$ and $t_i=2013.96\pm0.02\ year$, with $\text{BIC}=174.1$ , while for the Bass model $N=42.1\pm0.4$, $q=2.0\pm0.2\ year^{-1}$, $p=0.08\pm0.08\ year^{-1}$ and $t_i=2013.93\pm0.05\ year$, with $\text{BIC}=172.8$ (we do not show in the figure the fit with the Bass model since it overlaps the logistic). The fit provided by logistic and Bass are very similar but the former has less parameters, thus its BIC is smaller. In any case the fit is remarkably good considering the reduced statistics. Fitting the logistic model with data from $1$st to $42$th installations gives $N=43.7\pm0.7$, $q=1.69\pm0.08\ year^{-1}$ and $t_i=2014.00\pm0.04\ year$, provides also a good fit although not as accurate as the one before, judging it by the value $\text{BIC}=2793.5$.

Altogether, the overall evolution shows two successive stages of logistic grow: the first one with an inflection point in 2013 and an small saturation level that was reached in late 2015, and a second one starting in June 2016 with an inflection point reached in 2023 and characterized by a much larger saturation level (market capacity). In general these changes in the effective market capacity are associated to substantial legislative or technological changes~\cite{Bunea2020,other_entities}.

In the case study addressed here, the 2015 slowdown can be related to legislation changes, specifically the so-called \textit{impuesto al sol}. Implemented in October of 2015~\cite{RD2015} and repealed in October of 2018~\cite{RD2018}, it was a tax on self-consumption for installations larger than 10 kW connected to the grid. While justified as a toll to cover the cost of using the grid, it taxed all the generated energy, even if self-consumed never injected on the grid.
Fig.~\ref{fig_early_evolution} shows the number of searches of \textit{"impuesto al sol"} in Google~\cite{trends_impost}. The number of searches is relevant from March 2015 reflecting a public debate that started started several months before the law approval (already in July 2025 several political parties signed an agreement to revoke it if approved~\cite{manifesto_impuesto_sol_ER, manifesto_impuesto_sol_SER}). Reportedly an effect of this measure was to discourage new self-consumption installations~\cite{efecto_impuesto_sol}.
According to number of registers in RAC, the fraction of installations above $10$~kW decreased from $71.4\%$ before the tax to $17.6\%$ during its application. 

Remarkably, the tax also appears to have discouraged installations below $10$~kW, even though these systems were formally exempt from it. This effect is visible in the period from September 2015 to June 2016, during which virtually no new installations were registered. A plausible explanation is the widespread confusion and misinformation surrounding the implementation of the policy, which may have deterred potential adopters until the regulatory framework became clearer for households and installations began to increase again. A further peak in Google searches for ``\textit{impuesto al sol}” occurred when the tax was eventually repealed; however, by that time residential installations had already begun to grow rapidly and were largely unaffected by the repeal.

\section{Spatial analysis and correlations}
\label{Spatial_analysis}

The inset of Fig.~\ref{fig_early_evolution} shows the spatial distribution of self-consumption installations as of June 2016, reflecting the early stages of adoption. The installations are clearly not uniformly distributed across the territory, nor do they appear to have expanded outward from a single spatial origin. Instead, they are widely dispersed, with several distinct clustering areas. In particular, the industrial region surrounding Barcelona concentrates approximately one third of installations larger than $10$~kW, while roughly half of the installations below $10$~kW are located along the Barcelona/Manresa corridor.

\begin{figure}
\includegraphics[width=\columnwidth]{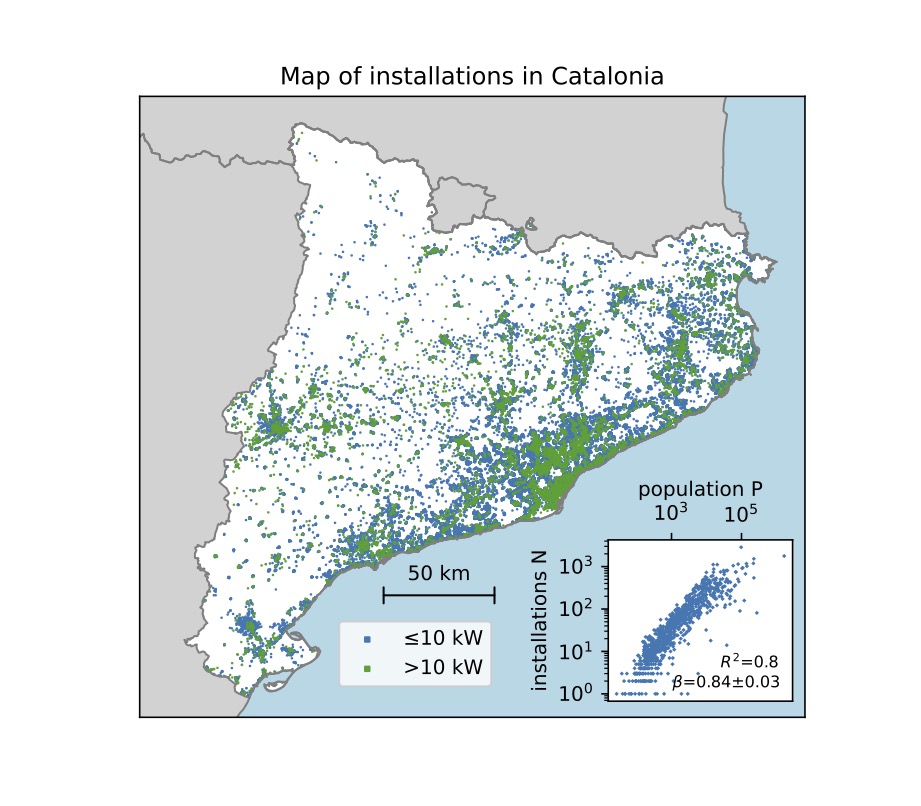}
\caption{Map of PV self-consumption installations disaggregated in smaller (blue) and larger (green) than $10$~kW by 31st of May, 2024. The inset shows the scatter plot of the number of installations smaller than 10 kW vs population per municipality in double logarithmic scale, with the corresponding Pearson's $R^2$ coefficient and slope.
}
\label{map_installations}
\end{figure}

We represented as an animated GIF the spatiotemporal evolution of self-consumption installations form 2013 to May 2024, disaggregated in residential (smaller than $10$~kW, in blue) and non-residential (larger than $10$~kW, in green) \cite{diffusion_gif}. The cumulative installations at the end of the period are shown in Fig.~\ref{map_installations}. While residential and non residential distributions are far from being homogeneous, the former is more spread out. The spatial distribution of the residential installations, to a large extend, can be interpreted as associated to population. This can be analyzed by considering the data for the $947$ municipalities into which Catalonia is divided. The inset shows the number of residential installations by municipality versus the population in double logarithmic scale. In this scale the relationship is close to being linear with a Pearson correlation value $R^2=0.8$ and a slope of $\beta=0.84\pm0.03$. Therefore, to a good extend, the distribution of installations follows a power law of the distribution of people. Power laws do not introduce characteristic scales, thus the characteristic spatial scales of the household installations are given by those of the population. 
Since the exponent is smaller than $1$ the number of installations grows slower than the population implying a saturation effect in largely populated municipalities. There are a few outliers, the most relevant ones being municipalities in the metropolitan area of Barcelona or in the Pyrenees.

Nevertheless, as we will show, factors beyond population also play an important role in shaping adoption. First, we analyze the temporal evolution at the level of \textit{comarques}, providing a finer spatial perspective than in the previous section. Second, we examine the influence of additional variables by considering the accumulated number of installations per capita as of May 2024 at the municipal level, the smallest administrative spatial scale. Finally, we investigate spatiotemporal correlations within the residential installations dataset.

\subsection{Temporal evolution by \it{comarques}}

\begin{figure}
\includegraphics[width=\columnwidth]{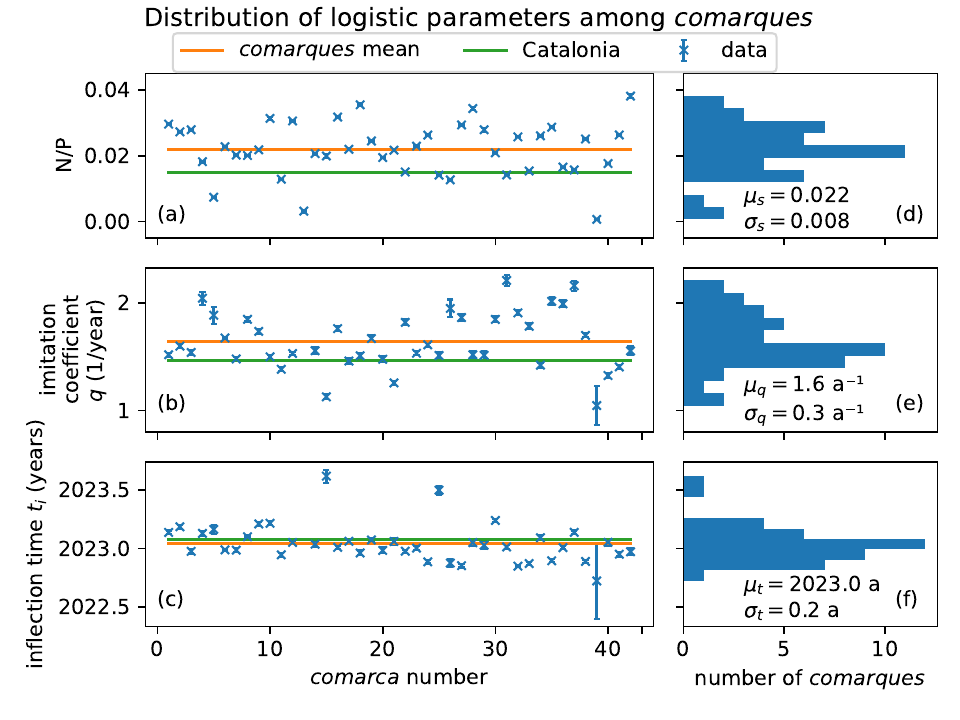}
\caption{Panels in the left column (a-c) show the parameter estimates for the logistic fit (saturation number normalized to population (N/P), imitation coefficient $q$, and inflection time $t_{\rm I}$ for each \textit{comarca} (blue). The values of $q$ and $t_{\rm I}$ for \textit{el Barcelon\`es} are outside the range of the panels. The orange horizontal lines show the averages of the values obtained for the different \textit{comarques} and the green ones the values for the overall fit with all data for Catalonia. Panels in the right column (d-f) display the histograms of the parameter distributions.} 
\label{figure6}
\end{figure}

In order to see if the temporal evolution observed before is similar at smaller spatial scales, we disaggregated the data for the $42$ \textit{comarques}, areas of 146--1784 km$^2$ into which Catalonia is administratively divided. The logistic model fits well the growth of residential installations for the majority of comarques with the exception of \textit{el Barcelon\`es}, the most densely populated one (see the supplementary material \ref{supplementary:fits_comarques} for fits and a table with parameter estimates). The parameter estimates of the fits are plotted in the left column of Fig.~\ref{figure6}. Given the diversity of the population of the \textit{comarques}, rather than the saturation level $N$ we plot its ratio to the population $P$ (top panel). The orange horizontal lines correspond to the average over all \textit{comarques}, while the green ones to the overall fit discussed in the previous sections. The histograms on the right column show the parameters distribution. All of them are bell-shaped with a peak close to the mean and have a relatively small standard deviation. Also the standard deviation of the inflection time is much smaller than the timescale of the growth, $q^{-1}$. These result can be seen as a validation of the logistic model at the scale of \textit{comarques}, and at the same time a hint to the existence of spatial heterogeneity, as will be studied in subsection~\ref{spatial_distr}.

\subsection{Spatial distribution of residential PV installations} \label{spatial_distr}

To further analyze spatial inhomogeneities beyond those directly given by population density we now consider the number of residential PV installations per capita disaggregated by municipalities. Here we just consider the cumulative number of installations at the end of the period (May 2024) disregarding the date of installation. The result is plotted Fig.~\ref{figure7} (see the supplementary material \ref{supplementary:installations_per_building} for a version of this figure using the number of buildings instead of the population).

The reduced number of installations in the northwestern municipalities can be understood, at least in part, from the fact that they correspond to the Pyrenees area, in which the number of sunshine hours is relatively small (see inset). There are $22$ ($2.3\%$) municipalities with no residential installations. Except for a singular case discussed later, the others are either located in the Pyrenees or have less than $300$ inhabitants. Leaving that aside, the most relevant result is that installations per capita are larger in the rural areas rather than in the most populated areas (cities and seaside municipalities). In fact, installations per capita are particularly low for the most density populated municipalities such as those belonging to \textit{el Barcelon\`es} (see the supplementary material \ref{supplementary:bcn_neighbourhoods} for details about this \textit{comarca}).

In Fig.~\ref{figure8} we plot for each municipality the density of installations versus the urban population density (calculated as the total population of the municipality divided by the urban land), the average household income (measured by the GDHI), and the fraction of buildings under 3 stories. Each panel also shows the value of the Pearson's correlation coefficient and the p-value. Given the low p-values the correlation values obtained are significant. The correlation correlation with household income is positive but is extremely weak, $R^{2}=0.08$, therefore it does not play a relevant role in adoption. More relevant is the negative correlation with the urban population density ($R^{2}=0.2$). The most relevant correlation, however, is the moderate and positive one with the fraction of buildings under 3 stories, with $R^{2}=0.5$. The fact that is considerably stronger than the others indicates that the housing type affects directly the decision to install. Multi-apartment buildings have a reduced roof surface per dwelling. As building density increases, electricity demand grows faster than usable solar surface, while shading from surrounding structures and rooftop obstacles further reduces effective generation capacity and economic performance~\cite{fuster2021innovative,Saez2023}. Besides, since the rooftop area is usually shared by many households it is necessary an agreement between neighbors and additional regulatory requirements associated with collective self-consumption schemes apply~\cite{autoconsumo_espana}. Furthermore some buildings may also be subject to aesthetic or heritage restrictions which handicap PV installations~\cite{Lucchi2022}. A singular case in this respect is Badia del Vall\`es, a municipality of $0,9$ Km$^2$ with $5400$ apartments in buildings $8$ to $12$ stories high. This municipality started as social housing which implies that additional permits are required for PV installations. Besides, there are no municipal tax reductions. As a result, by May 2024 this was the only municipality in Catalonia larger than $1000$ people with no registered self-consumption installations. 

\begin{figure}
\includegraphics[width=\columnwidth]{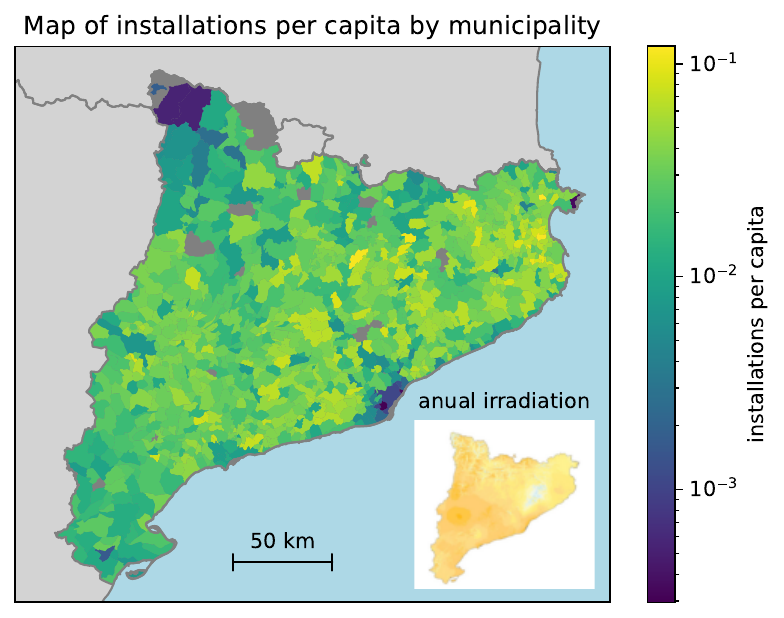}
\caption{Installations per capita by municipality with a logarithmic color scale. Municipalities with no installations are displayed in dark gray. In the inset, map of the annually averaged solar irradiation.}
\label{figure7}
\end{figure}

\begin{figure}
\includegraphics[width=\columnwidth]{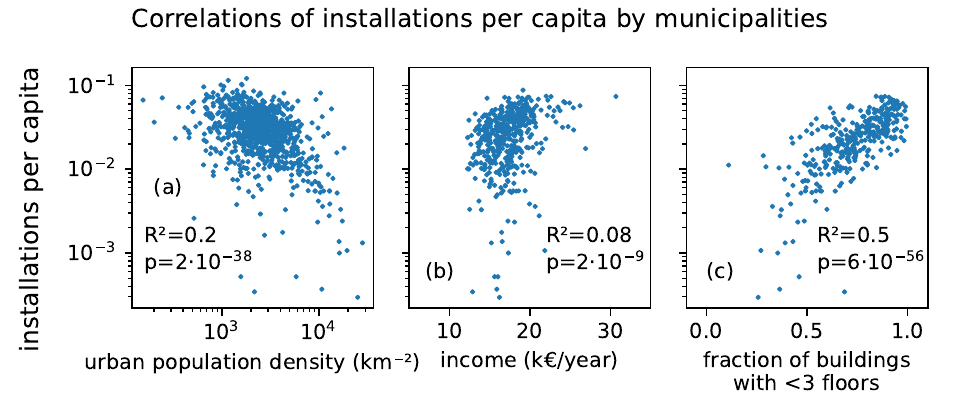}
\caption{Scatter plots of the logarithm of the number of installations per capita as a function of (a) the logarithm of the urban population density (taking into account only urban soil), (b) the GDHI (a measure of the income), and (c) the fraction of buildings less than 3 stories high, all by municipality. The Pearson's correlation coefficient and its p-values are indicated for each case. The number of municipalities considered vary depending on the available data.}
\label{figure8}
\end{figure}

\subsection{Temporal correlations between installations}

\begin{figure}
\includegraphics[width=\columnwidth]{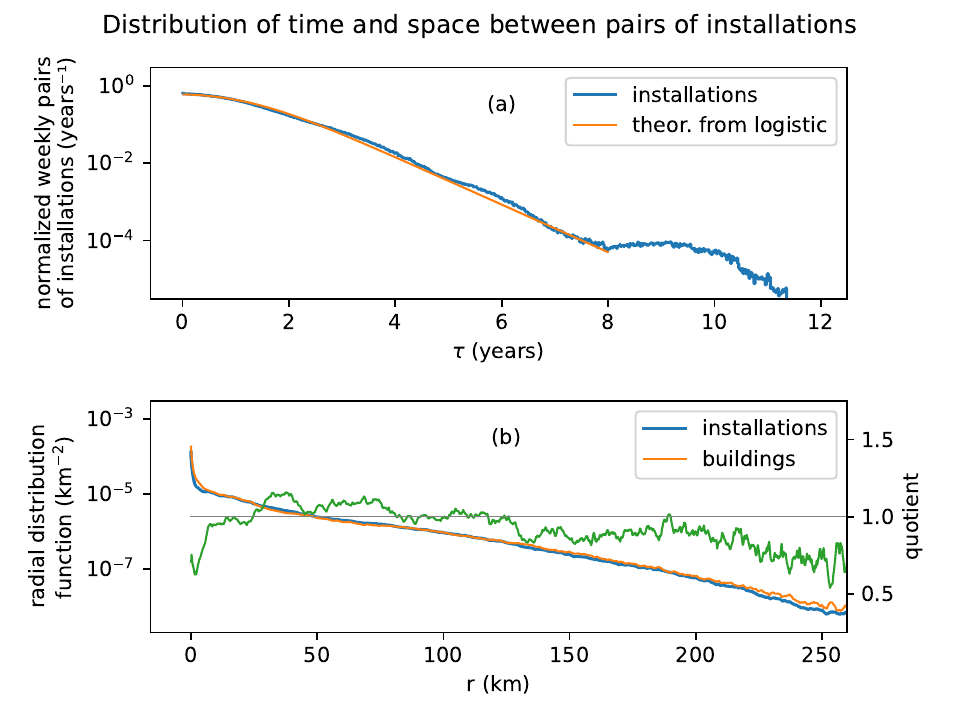}
\caption{(a) Temporal pair distribution $C(\tau)$ of installations separated by a time interval $\tau$ (blue) and theoretical approximation~\ref{eq:C_tau_final} (orange). (b) Spatial pair distribution $\rho(r)$ of installations (blue) and buildings (orange) separated by a distance $r$, and their ratio (green).
}
\label{fig_correlations}
\end{figure}

We analyze the temporal pair distribution $C(\tau)$, defined as the probability density that two installations are separated by a time interval $\tau$ (see \ref{appendix:ctau} for a precise definition and details on its numerical estimation). 

Under the assumption that the adoption dynamics follow a logistic growth, it is possible to derive an approximate analytical expression for the shape of $C(\tau)$ (see Eq.~(\ref{eq:C_tau_final}) in \ref{appendix:ctau}). As shown in panel (a) of Fig.~\ref{fig_correlations}, the analytical approximation provides a good description of the empirical $C(\tau)$ for $\tau < 8$ years. In this regime, the pair statistics are dominated by installation events occurring within the main growth phase (after June 2016), during which the dynamics are well captured by a single logistic process. The agreement is particularly strong for $\tau < 3$ years, where most contributing pairs belong to this regime.

For larger time separations, $\tau > 8$ years, $C(\tau)$ becomes increasingly dominated by pairs formed by one installation from the early phase (before June 2016) and another from the main growth phase (after June 2016). In this case, the assumption of a single logistic process with fixed parameters is no longer valid, as the two periods are characterized by different growth dynamics. As a result, the analytical approximation derived under homogeneous logistic growth fails to accurately reproduce the empirical distribution in this regime.

\subsection{Spatial correlations between installations}

We now focus on the spatial pair distribution $\rho(r)$, defined as the probability density that two installations are separated by a distance $r$, irrespective of their installation time (see \ref{appendix:rhor} for a precise definition and details of its numerical estimation).

As shown in panel (b) of Fig.~\ref{fig_correlations}, the spatial pair distribution decays sharply at short distances ($r<10$~km). This is followed by an approximately exponential decay over a broad range of distances. Because of the finite spatial extent of Catalonia, the distribution becomes dominated by finite-size effects at very large distances ($r > 250$~km), and beyond $300$~km it rapidly approaches zero.

Since residential PV installations are closely associated with buildings, we compare this result with the distribution of pairs of buildings separated by a given distance (orange line). This reference distribution was computed using data on the spatial distribution of residential buildings in Catalonia (approximately one million)~\cite{corral2022finite, corral_private_comm}.

While the two distributions are broadly similar, some systematic differences can be observed. The radial distribution of installation pairs is lower than that of buildings at both short ($r < 30$ km) and large ($r > 130$ km) distances, whereas the opposite occurs at intermediate distances (see green line). The smaller value at short distances can be explained by the lower density of installations in highly populated metropolitan areas, as discussed previously. The lower number of installation pairs at large distances partly reflects the relatively small number of installations in the Pyrenees and southern Catalonia, which are the regions located farthest from the main concentration of installations. Given these deficits at short and long distances, the normalization of the distributions to unit area results in a comparatively higher value for installation pairs at intermediate distances.

\section{Conclusions}
\label{Conclusions}

We have analyzed the spatiotemporal evolution of PV residential installations on the framework of diffusion of innovation models, characterizing the effects of social interactions as well as demographic, urban and socioeconomic environments. Using the data for PV self-consumption installations in Catalonia, we show that the nominal capacity can be used as a proxy to to classify residential and non-residential installations in data sets in which this information is not provided. 

We identify two distinct stages of logistic growth. The first phase exhibits an inflection point around 2013 and reaches a relatively low saturation level by late 2015. This is followed by a second phase beginning in June 2016, with an inflection point in early 2023 and a substantially larger market capacity (approximately $118,600$ installations). While changes in market capacity are typically associated with legislative or technological shifts~\cite{Bunea2020, other_entities}, in this case they appear to be largely mediated by social perception. The introduction of a new tax—despite not applying to residential installations—generated widespread concern and led to a stagnation in installations in late 2015.

We also find that, in both growth phases, including the innovation parameter of the Bass model does not improve the fit compared to the logistic model. At smaller spatial scales, such as the level of \textit{comarques}, the logistic model continues to provide a good description of the dynamics. With the exception of the most densely populated \textit{comarca}, parameter estimates are consistent across regions, indicating that the observed patterns are robust over a wide range of spatial scales.

We also introduce a method, based on a reinterpretation of the role played by the function $\chi(t)$ in the generalized Bass model, to empirically quantify temporal deviations between the data and the model, thereby enabling the analysis of external influences. Focusing on the later—and dominant—growth stage, we find that the most significant external drivers are subsidies provided by the Catalan Government for PV installations and the perceived cost of electricity. Notably, social perception of electricity prices appears to play a more important role than the actual price itself.

Considering the spatial distribution of residential installations, we find that it closely follows the distribution of the population, with a power-law relationship characterized by a slope of $0.84 \pm 0.03$. Since power laws do not introduce intrinsic spatial scales, this result indicates that the characteristic spatial scales of residential PV adoption are largely determined by those of the population distribution. After accounting for population size by analyzing installations on a per capita basis, lower adoption levels are observed in mountainous municipalities (particularly in the Pyrenees), in very small municipalities with fewer than 300 inhabitants, and in the most densely populated areas. In particular, we identify a clear correlation between PV adoption and the fraction of buildings with fewer than three stories.

Finally, we show that the temporal pair distribution $C(\tau)$ of residential installations is well described by the analytical expression derived under the assumption of logistic growth, at least within the regime where a single growth process dominates. In addition, the spatial pair distribution $\rho(r)$ closely follows that of residential buildings, indicating that the spatial organization of PV adoption largely reflects the underlying structure of the built environment rather than additional spatial interaction effects.

The analysis presented here indicates that the growth of residential self-consumption PV installations in Catalonia is approaching saturation. However, the predicted saturation level corresponds to only about $12\%$ of the total number of residential buildings in the region. This indicates that the slowdown is not driven by direct physical constraints, but rather by the prevailing regulatory and socio-economic context.

The fact that two successive growth phases have already occurred suggests that additional expansion could still take place if significant legislative or technological changes were introduced and if these changes were clearly perceived by the public. In particular, urban areas characterized by multi-apartment buildings exhibit the lowest number of installations per capita, for the reasons discussed in Section~\ref{Spatial_analysis}, and therefore represent the largest potential for future adoption. Facilitating collective installations, by streamlining administrative procedures or providing targeted incentives, would be clearly beneficial in this regard, since collective installations represented only 1.2\% of the self-consumption PV installations in Catalunya in 2023~\cite{obercat}. Nevertheless, as emphasized earlier, the effectiveness of such measures depends not only on their implementation but also on their visibility and public awareness.

From a theoretical point of view, the models considered describe the overall growth disregarding specific interactions between individuals. For many social processes this is far from being realistic~\cite{gleeson2012accuracy, Manshadi2020,manshadi2016generalized}, so it is remarkable that it leads to good fits. This may come from the interaction network having small-world properties, providing an overall good connectivity with a reduced number of links~\cite{watts1998collective}. In fact some surveys found a bigger relevance of close contacts (like friends or family) which, compared to neighbors, are of a longer range~\cite{rai2013effective,palm2017peer}. Besides, in this context, weak ties (interactions with acquaintances not sharing the same close friends) become collectively relevant for diffusion of information, innovations, or other concepts~\cite{granovetter1983strength}.
These explanations are coherent with the spatial homogeneity of the evolution (section~\ref{spatial_distr}), since at least some long range interactions are needed to explain why there is practically no spatiotemporal clustering (the preference of new installations to be located near existent ones). 

Finally, we remark that although our work focuses on Catalonia as a case study, the analytical framework developed here for the diffusion of residential PV installations can be readily applied to other regions, provided that sufficiently detailed data are available. We expect similar qualitative patterns to emerge, including the presence of successive growth phases shaped by the legislative and fiscal environment, lower installation rates per capita in densely populated urban areas, and the dominant role of socially perceived changes—rather than policy changes alone—in driving adoption dynamics.

\section*{Acknowledgements}
Partial financial support has been received from Ajuntament de Barcelona and Fundació “la Caixa” under the project \textit{De l’euro al joule} (Eur2J), the Agencia Estatal de Investigaci\'on (AEI, MCI, Spain)
MCIN/AEI/10.13039/501100011033 and Fondo Europeo de Desarrollo Regional (FEDER, UE) under Projects  PID2024-157493NB-C21, PID2024-157493NB-C22 and the Mar{\'\i}a de Maeztu Program for units of Excellence in R\&D, grant CEX2021-001164-M.

\bibliography{biblio_v2}

\begin{appendix}

\section{Distribution of pairs of installations separated by a time interval in the logistic model}
\subsection{Temporal correlations}
\label{appendix:ctau}
We define the temporal pair distribution $C(\tau)$ such that $C(\tau) \Delta \tau$ represents the fraction of pairs of installations whose difference in installation times lies within the interval $(\tau, \tau +\Delta\tau)$. The distribution is normalized over the observation window, so that
\begin{equation}
\label{eq:norm}
\int_{0}^{t_{\rm f}-t_0} C(\tau)\, d\tau = 1,
\end{equation}
where $t_0$ and $t_{\rm f}$ denote the initial and final observation times, respectively. For the numerical evaluation of this quantity using the dataset, we generate a histogram of time differences using all pairs of installations and a binning size $\Delta \tau=1~\text{week}$, and normalize it afterwards according to Eq.~\eqref{eq:norm}.

Assuming statistical independence between installation events at different times, the temporal pair distribution can be approximated as the convolution
\begin{equation}
C(\tau) = 2\int_{t_0}^{\tf-\tau} x(t)x(t+\tau)\, dt,
\label{eq:temporal_pair_distribution}
\end{equation}
where $x(t) = \frac{1}{\nf-n_0}\frac{dn(t)}{dt}$ is such that $x(t)dt$ represents the fraction of new installations occurring within the time interval $(t, t+dt)$, divided by the difference between the cumulative installations at the final $\nf = n(\tf)$ and initial $n_0=n(t_0)$ times.

For the logistic model, we use Eq.~\eqref{eq:S_logistic}, to obtain
\begin{align}
 x(t)&
 = \frac{qN}{n_\text{f}-n_0}\frac{\exp(-q(t-t_{\rm I}))}{(1+\exp(-q(t-t_{\rm I})))^{2}} .
 \label{eq:growth_logistic}
\end{align}

Introducing Eq.~(\ref{eq:growth_logistic}) in (\ref{eq:temporal_pair_distribution}), and defining \makebox{$u_\tau\equiv \exp(q\tau)$}, we have:
\begin{multline}
C(\tau)=\frac{2 q N^2 u_\tau}{(n_\text{f}-n_0)^2(1-u_\tau)^2} \left[\frac{1+u_\tau}{1-u_\tau}\ln\left(\frac{1+e^{-q(t-t_\text{I})}}{u_\tau+ e^{-q(t-t_\text{I})}}\right)\right. 
    \\
\left. \left. +\frac{1}{e^{q(t-t_\text{I})}+1} + \frac{1}{u_\tau e^{q(t-t_\text{I})}+1} \right] \right |_{t_0}^{t_\text{f}-\tau}. 
\label{eq:C_tau}
\end{multline}
The logarithm term can be rewritten as
\begin{align}
    &\left . \ln \left(\frac{1+e^{-q(t-t_\text{I})}}{u_\tau+ e^{-q(t-t_\text{I})}}\right) \right |_{t_0}^{t_\text{f}-\tau}  = \ln\left( \frac{1+u_\tau e^{-q(t_\text{f}-t_\text{I})}}{u_\tau+u_\tau e^{-q(t_\text{f}-t_\text{I})}} \frac{u_\tau+e^{-q(t_0-t_{\text{I}})}}{1+e^{-q(t_0-t_\text{I})}}\right) \nonumber \\
    &= \ln \left( \frac{n_\text{f}(1-u_\tau) + u_\tau N}{u_\tau N} \frac{(u_\tau-1)n_0 + N}{N}\right) \nonumber \\
    &\approx  \ln \left( \frac{n_\text{f}(1-u_\tau) + u_\tau N}{u_\tau N} \right) ,
    \label{eq:C_log_term}
\end{align}
where in the last approximation we have used $n_0 \ll N$. Similarly
\begin{multline}
  \left . \frac{1}{e^{q(t-t_\text{I})}+1}  \right |_{t_0}^{t_\text{f}-\tau}  =
  \frac{u_\tau}{e^{q(t_\text{f}-t_\text{I})}+u_\tau} - \frac{1}{e^{q(t_0-t_\text{I})}+1} \\ 
  =  \frac{u_\tau(N-n_\text{f})}{n_\text{f}(1-u_\tau)+u_\tau N}- \frac{N-n_0}{N}
  \approx  - \frac{n_\text{f}}{n_\text{f}(1-u_\tau)+u_\tau N} ,
  \label{eq:C_frac1}
\end{multline}
and
\begin{multline}
\left. \frac{1}{u_\tau e^{q(t-t_\text{I})}+1} \right |_{t_0}^{t_\text{f}-\tau} = \frac{1}{e^{q(t_\text{f}-t_\text{I})}+1} - \frac{1}{u_\tau e^{q(t_0-t_\text{I})}+1}  \\
= \frac{N-n_\text{f}}{N} - \frac{N-n_0}{(u_\tau-1)n_0+N} \approx - \frac{n_\text{f}}{N}. 
\label{eq:C_frac2}
\end{multline}
Introducing (\ref{eq:C_log_term}) - (\ref{eq:C_frac2}) in (\ref{eq:C_tau}) leads to

\begin{multline}
C(\tau) \approx \frac{2 q N u_\tau}{n_\text{f}(1-u_\tau)^2} \left[\frac{1+u_\tau}{1-u_\tau} \frac{N}{n_\text{f}}\ln \left( \frac{n_\text{f}(1-u_\tau) + u_\tau N}{u_\tau N} \right)\right. 
    \\
\left. - \frac{N}{n_\text{f}(1-u_\tau)+u_\tau N} - 1  \right],
\label{eq:C_tau_final}
\end{multline}
which is the function fitted in Fig.~\ref{fig_correlations}a.

\subsection{Spatial correlations}
\label{appendix:rhor}

We define the spatial pair distribution $\rho(r)$ such that $\rho(r)\, r\, \Delta r$ represents the fraction of pairs of installations whose separation distance lies within the interval $(r, r+\Delta r)$. The factor $r$ accounts for the fact that, in two-dimensional space, the anndifferential area is $dA(r) = 2\pi r\, dr$. The normalization condition is therefore
\begin{equation}
\label{eq:normrho}
\int_0^{\infty} \rho(r)\, dA(r) = \int_0^{\infty} \rho(r)\, 2\pi r\, dr = 1,
\end{equation}
so that a spatially uniform distribution yields a constant $\rho(r)$.

For the numerical evaluation, we compute all pairwise distances between installations present at the final observation time $\tf$ and construct a histogram using a bin width $\Delta r = 100~\text{m}$. The resulting distribution is then normalized according to Eq.~\eqref{eq:normrho}.

\end{appendix}
\clearpage

\renewcommand{\thesection}{SM\arabic{section}}
\setcounter{section}{0}
\renewcommand{\thefigure}{SM\arabic{section}-\arabic{figure}}
\setcounter{figure}{0}
\renewcommand{\thetable}{SM\arabic{table}}
\setcounter{table}{0}

\begin{center}{\Large Supplementary Material}\end{center}

\section{Characteristics and processing of the Self-consumption Register of Catalonia} \label{supplementary:dataset_info}

The Self-consumption Register of Catalonia has been in use since its establishment following the publication of Royal Decree 15/2018 \cite{RD2018} in October 2018, and is updated periodically, typically every few months. According to Article 2 of Royal Decree 244/2019 \cite{RD2019}, the register includes all self-consumption energy installations, with the exception of isolated systems and ``generation units used exclusively in the event of a disruption in the electricity supply from the grid.''

Some isolated installations may exist that are neither connected to the grid nor benefit from subsidies or tax incentives. However, their number is expected to be small, as grid connection with surplus compensation is generally the more economically advantageous option, and the vast majority of the population resides in areas with access to the electricity network. On this basis, we assume that the register captures most installations and provides a representative picture of the state of photovoltaic self-consumption in Catalonia.

Two dates are recorded in the dataset: I\_DAT\_INSCRIP\_RAC, corresponding to the date of registration, and I\_DAT\_PEM, indicating when the installation began operation. While these dates usually coincide, they may differ for older installations, as the registration date is constrained by the creation of the register itself. In our analysis, we use the latter, since it more accurately reflects the moment when agents decided to undertake the installation and allows us to study the evolution prior to RD15/2018 \cite{RD2018}. These earlier installations, developed under previous regulatory frameworks, were subsequently required to comply with the current one, as established in the \textit{Disposici\'on transitoria primera} of RD244/2019 \cite{RD2019}. As a result, their registration date corresponds to the time of regulatory adaptation, whereas the date of entry into operation reflects their actual commissioning at an earlier point in time.

Regarding the technology, we restrict our analysis to photovoltaic installations, as they account for virtually the entire dataset (99.96\% of the total). From an agent-based modeling perspective, this choice also avoids introducing heterogeneity associated with mixing different technologies.

Concerning the \textit{comarques}, the administrative region of \textit{el Lluçanès} was created on May 3, 2023. However, to ensure consistency over time, we use the previous \textit{comarques} division for all installations in our analysis.

Finally, with respect to spatial coordinates, most records are provided in UTM format—a projection that preserves distances locally—with a typical resolution of one meter. A smaller subset of data is given in geographic coordinate system (GCS) format (latitude and longitude) and was therefore converted to UTM. In addition, various inconsistencies and errors were corrected, as discussed in the article, and some coordinates were discarded when they could not be reliably interpreted within any known coordinate system.

\section{Robustness of the saturation estimations}
\label{supplementary:robustness}

\begin{figure}[h]
\begin{center}
\includegraphics[width=\columnwidth]{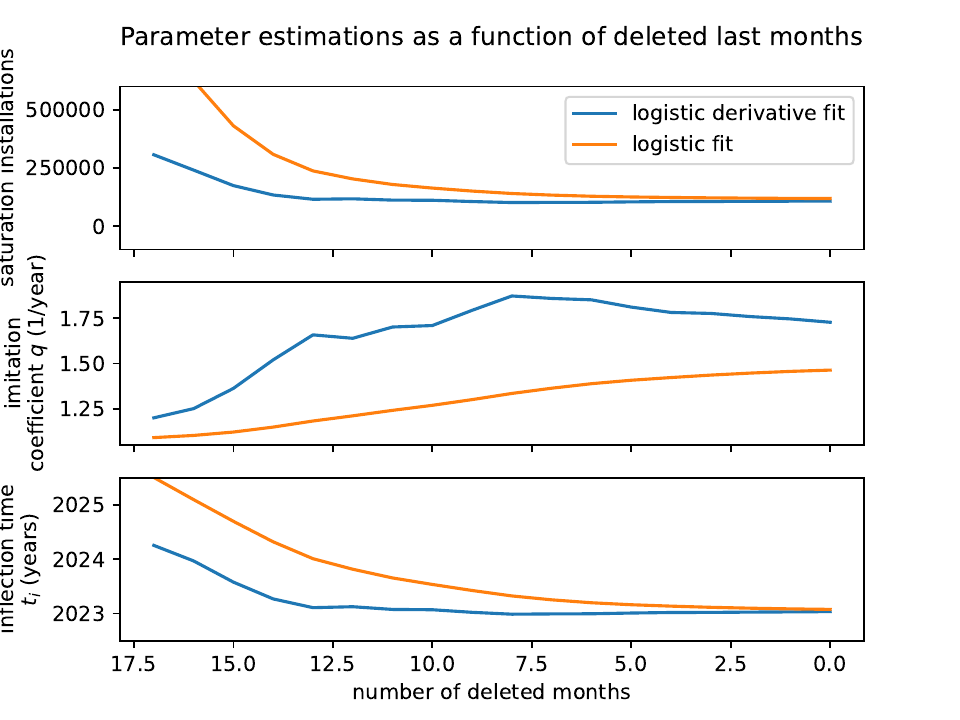}
\end{center}
\caption{Estimates of the parameters of the logistic function (installations at saturation, imitation coefficient and inflection time) as a function of the number of deleted recent months, with both the logistic function and its derivative.}
\label{sfigure1}
\end{figure}

To assess the robustness of the logistic fit estimations, we recomputed the model parameters while progressively removing the most recent months of data. The results, shown in Fig.~\ref{sfigure1}, indicate convergence after the peak, which occurs approximately 16 months before the end of the dataset (May 2024). This behavior is consistent with previous findings on the robustness of diffusion models \cite{van1997bias}, which highlight the sensitivity of parameter estimates to the temporal range of the data, particularly when observations end near or before the inflection point. However, our results differ from that study in the direction of the bias, possibly due to differences in the modeling approach (logistic versus Bass) or in the estimation method (Trust Region Reflective algorithm versus nonlinear least squares). It is worth mentioning that estimates derived from the cumulative curve and its derivative exhibit different convergence patterns: the cumulative fit, due to its smoothness, yields more stable but slower-converging estimates, whereas the derivative-based fit, although noisier, appears to approach the converged values more rapidly.

\section{Testing of alternative mechanisms of interaction}
\label{supplementary:mechanisms}

\begin{figure}[h]
\begin{center}
\includegraphics[width=\columnwidth]{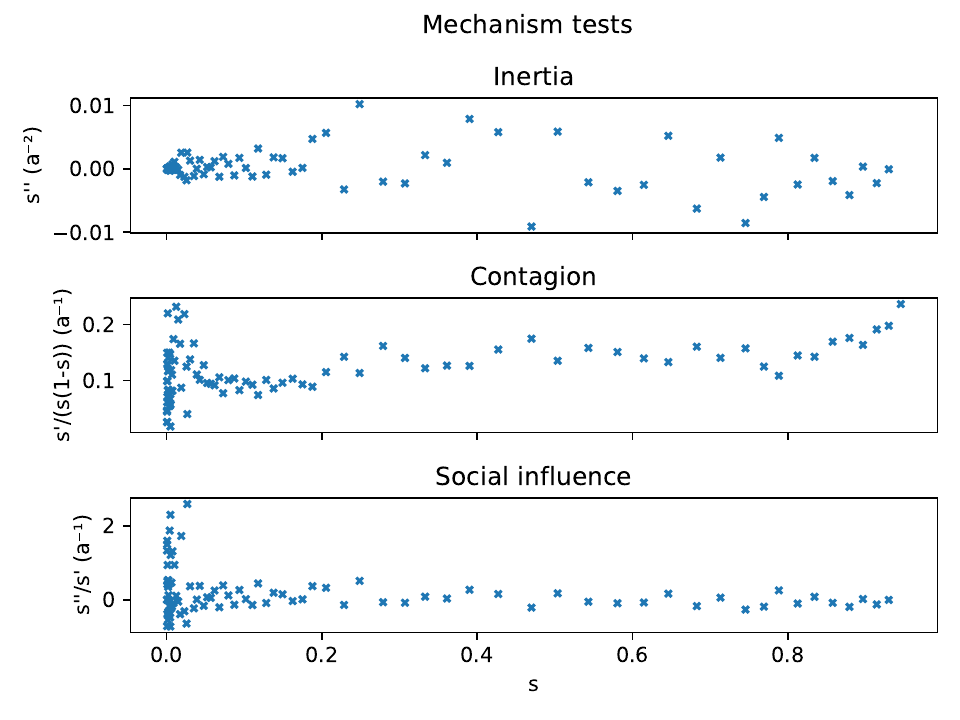}
\end{center}
\caption{Different scatter plots involving the cumulative fraction of installations $s(t)$ and its first and second derivatives (discretized, with monthly resolution), testing different models: $s'$ as a function of $s$ for inertia, $s'/(s(1-s)$ as a function of $s$ for contagion, and $s''/s'$ as a function of $s$ for social influence.}
\label{sfigure2}
\end{figure}

Following the methodology proposed by Young \cite{young2009innovation}, we tested whether specific properties—characteristic of certain diffusion models—could be identified in the installation time series, with the aim of supporting or ruling out alternative interaction mechanisms. The results are presented in Fig.~\ref{sfigure2}.

The first model, inertia, is obtained by assuming that the adoption rate $\omega(s)$ (where $s(t)\equiv n(t)/N$) is constant, i.e., independent of the state of the system. In the Bass framework, this corresponds to the innovation term. Under this assumption, the curvature of the adoption curve, $s''$, should be negative regardless of how this constant is distributed across the population. However, this behavior is not observed in the data: the second derivative fluctuates around zero, with approximately as many points above as below. This result does not rule out inertia as a contributing mechanism, but it indicates that it cannot be the sole driver of the observed dynamics.

The second model, contagion, is the one analyzed in the main paper. It encompasses mechanisms based on agent-to-agent diffusion of the innovation, such as the logistic model (corresponding to the imitative term in the Bass framework) and the non-symmetric responding logistic (NSRL), characterized by a non-linear adoption rate with dynamics given by $s' = q s^\delta (1 - s)$ \cite{easingwood1981nonsymmetric}. This class of mechanisms is compatible with the data. However, the diagnostic proposed by Young—based on the slope of $s'/(s(1-s))$—does not allow us to distinguish between Bass (negative slope), logistic (constant slope), and NSRL (positive slope) behaviors. The reason is that the empirical data are noisy, and the estimated slope is effectively indistinguishable from zero.

The third model corresponds to social influence driven by thresholds. The proposed test consists of examining whether the sign of $s''/s'$ at the early stages of diffusion remains stable over a period characterized by super-exponential growth. In our case, this quantity is highly unstable and frequently changes sign, which suggests that threshold-based social influence is unlikely to be the dominant mechanism driving the observed diffusion process.

\section{Fits of temporal evolution by \textit{comarques}}
\label{supplementary:fits_comarques}

\begin{figure}[h]
\includegraphics[width=\textwidth]{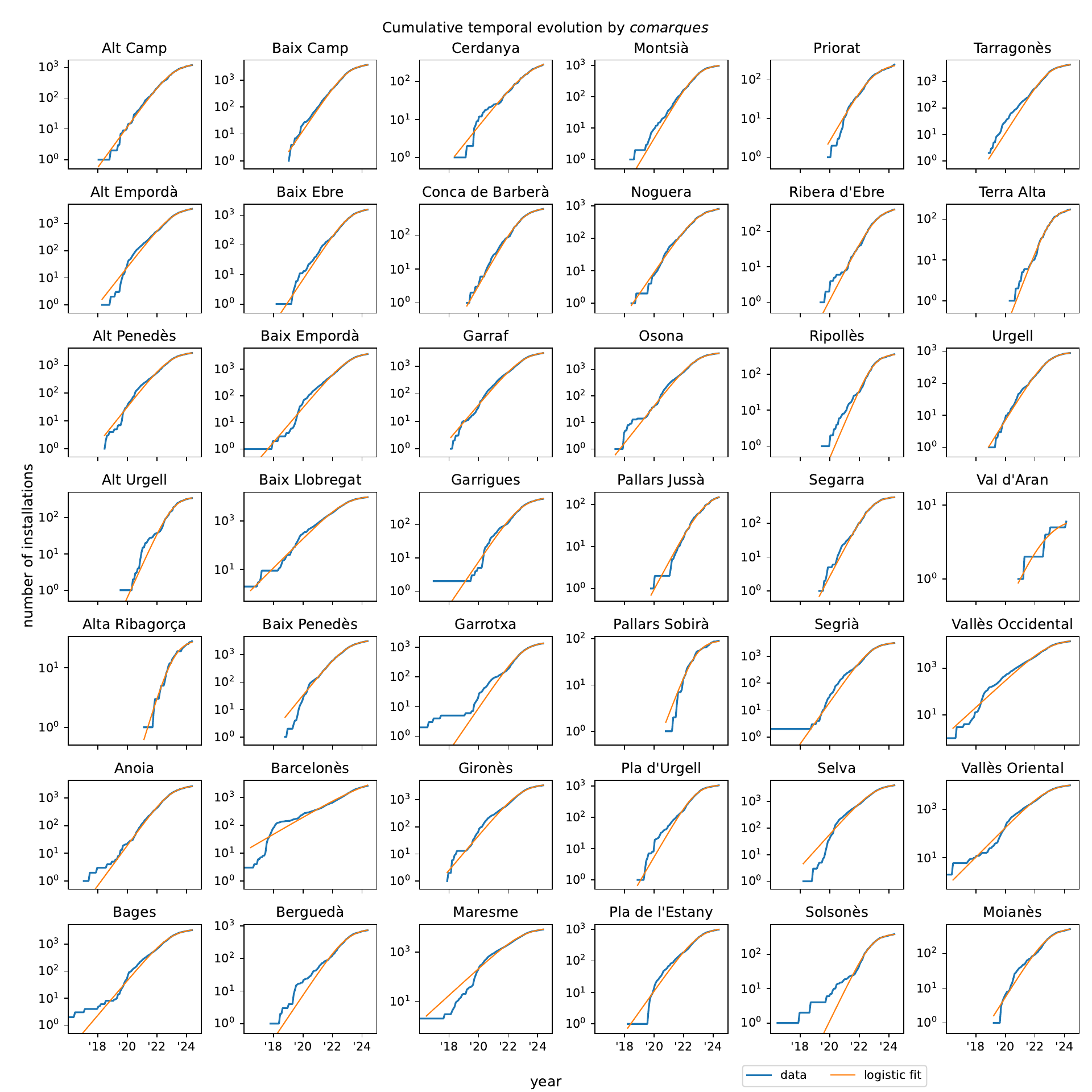}
\caption{Cumulative number of monthly installations as a function of time (blue) for each \textit{comarca}, together with the corresponding logistic fit (orange), shown on a logarithmic vertical scale. The fitting interval extends from the date of the first installation in each \textit{comarca}, or June 2016 (whichever is later), to May 2024.
}
\label{sfigure4}
\end{figure}

As discussed for Catalonia as a whole, the logistic model appears to be the most appropriate among the alternatives considered. We therefore also fitted this model separately for each of the \textit{comarques}. The resulting fits are shown in Fig.~\ref{sfigure4}, and the corresponding parameter estimates are reported in Table~\ref{stable1}. 

Visual inspection indicates that the overall behavior is broadly consistent across \textit{comarques}, although some variability is present. In general, the fits capture well the dynamics in recent years; however, at earlier stages, the logistic curves sometimes deviate from the data, both above and below. These discrepancies can be attributed to statistical fluctuations arising from the relatively small number of installations during the initial period.

One notable exception is the \textit{comarca} of \textit{el Barcelonès}, where the logistic model provides a poorer fit. In this case, the highly urbanized environment and the prevalence of apartment buildings act as barriers to installation, delaying the adoption process, as discussed in section \ref{spatial_distr}.

\onecolumn

\clearpage
\captionsetup[table]{width=0.9\linewidth}

\begin{longtable}{m{3.75cm} m{1.75cm} m{2.0cm} m{1.75cm} m{2.25cm} m{1.25cm}}
    \hline
    \textit{comarca} number and name & 2023 population & saturation density & $q\ (1/year)$ & $t_i\ (years)$ & 1st inst. \\
    \hline
    1\ \ Alt Camp & 46076 & 0.0296 (3) & 1.52 (2) & 2023.14 (2) & 01/2018 \\
    2\ \ Alt Empordà & 146766 & 0.0273 (3) & 1.60 (2) & 2023.18 (2) & 04/2018 \\
    3\ \ Alt Penedès & 112460 & 0.0279 (3) & 1.54 (2) & 2022.97 (2) & 06/2018 \\
    4\ \ Alt Urgell & 20762 & 0.0182 (3) & 2.04 (6) & 2023.13 (2) & 07/2019 \\
    5\ \ Alta Ribagorça & 4019 & 0.0074 (2) & 1.88 (8) & 2023.16 (4) & 02/2021 \\
    6\ \ Anoia & 126752 & 0.0228 (1) & 1.67 (1) & 2022.99 (1) & 01/2017 \\
    7\ \ Bages & 183265 & 0.0202 (2) & 1.48 (2) & 2022.99 (2) & 12/2013 \\
    8\ \ Baix Camp & 201647 & 0.0201 (1) & 1.85 (2) & 2023.10 (1) & 01/2019 \\
    9\ \ Baix Ebre & 81334 & 0.0218 (2) & 1.74 (2) & 2023.21 (1) & 03/2018 \\
    10 Baix Empordà & 141329 & 0.0313 (3) & 1.50 (2) & 2023.21 (2) & 01/2014 \\
    11 Baix Llobregat & 840572 & 0.0129 (1) & 1.38 (2) & 2022.94 (2) & 01/2013 \\
    12 Baix Penedès & 115701 & 0.0306 (2) & 1.53 (1) & 2023.05 (1) & 10/2018 \\
    13 Barcelonès & 2313975 & 0.0032 (4) & 0.71 (2) & 2025.12 (30) & 09/2008 \\
    14 Berguedà & 40618 & 0.0207 (3) & 1.56 (3) & 2023.04 (3) & 10/2017 \\
    15 Cerdanya & 19885 & 0.0199 (6) & 1.13 (2) & 2023.62 (6) & 05/2018 \\
    16 Conca de Barberà & 20480 & 0.0318 (3) & 1.76 (2) & 2023.01 (1) & 03/2019 \\
    17 Garraf & 159124 & 0.0220 (3) & 1.46 (2) & 2023.06 (2) & 02/2018 \\
    18 Garrigues & 18935 & 0.0355 (4) & 1.51 (3) & 2022.96 (2) & 12/2016 \\
    19 Garrotxa & 61363 & 0.0245 (3) & 1.67 (3) & 2023.07 (2) & 04/2014 \\
    20 Gironès & 201615 & 0.0195 (3) & 1.48 (3) & 2022.98 (2) & 11/2017 \\
    21 Maresme & 467398 & 0.0217 (3) & 1.26 (2) & 2023.06 (3) & 05/2002 \\
    22 Montsià & 70244 & 0.0151 (1) & 1.82 (3) & 2022.98 (1) & 05/2018 \\
    23 Noguera & 39567 & 0.0229 (2) & 1.53 (2) & 2023.00 (2) & 06/2018 \\
    24 Osona & 167506 & 0.0263 (2) & 1.61 (2) & 2022.89 (1) & 05/2017 \\
    25 Pallars Jussà & 13409 & 0.0141 (3) & 1.51 (3) & 2023.50 (4) & 10/2019 \\
    26 Pallars Sobirà & 7288 & 0.0127 (3) & 1.95 (8) & 2022.87 (3) & 10/2020 \\
    27 Pla d'Urgell & 37737 & 0.0294 (3) & 1.86 (4) & 2022.85 (2) & 11/2018 \\
    28 Pla de l'Estany & 33194 & 0.0344 (5) & 1.52 (3) & 2023.05 (3) & 03/2018 \\
    29 Priorat & 9360 & 0.0279 (6) & 1.51 (4) & 2023.03 (4) & 11/2019 \\
    30 Ribera d'Ebre & 22040 & 0.0209 (2) & 1.85 (3) & 2023.24 (2) & 05/2019 \\
    31 Ripollès & 25780 & 0.0142 (1) & 2.21 (5) & 2023.01 (1) & 06/2019 \\
    32 Segarra & 23938 & 0.0257 (2) & 1.91 (2) & 2022.85 (1) & 04/2019 \\
    33 Segrià & 215476 & 0.0154 (1) & 1.78 (3) & 2022.87 (1) & 04/2013 \\
    34 Selva & 182614 & 0.0260 (3) & 1.42 (2) & 2023.09 (2) & 03/2018 \\
    35 Solsonès & 13725 & 0.0287 (3) & 2.01 (4) & 2022.89 (1) & 06/2016 \\
    36 Tarragonès & 270237 & 0.0165 (1) & 1.99 (3) & 2023.01 (1) & 11/2018 \\
    37 Terra Alta & 11473 & 0.0157 (2) & 2.16 (5) & 2023.14 (2) & 04/2020 \\
    38 Urgell & 37960 & 0.0251 (2) & 1.70 (2) & 2022.89 (1) & 11/2018 \\
    39 Val d'Aran & 10496 & 0.0007 (1) & 1.05 (18) & 2022.72 (33) & 11/2020 \\
    40 Vallès Occidental & 949026 & 0.0176 (3) & 1.33 (2) & 2023.05 (3) & 10/2014 \\
    41 Vallès Oriental & 422149 & 0.0263 (3) & 1.41 (2) & 2022.95 (2) & 05/2013 \\
    42 Moianès & 14668 & 0.0381 (7) & 1.55 (4) & 2022.97 (3) & 03/2019 \\
    \hline
    \caption{Estimates of the logistic fit parameters by \textit{comarca}—including the saturation density (defined as the ratio between the saturation level and the population), the imitation coefficient $q$, and the inflection time—together with their associated uncertainties. The table also reports the month and year of the first installation, the \textit{comarca} identifier, and the population as of January 1, 2023. The fitting interval extends from the date of the first installation in each \textit{comarca}, or June 2016 (whichever is later), to May 2024.
}
    \label{stable1}
\end{longtable}

\twocolumn

\section{Map of the number of installations per building by municipalities}
\label{supplementary:installations_per_building}

\begin{figure}[h]
\begin{center}
    \includegraphics[width=\columnwidth]{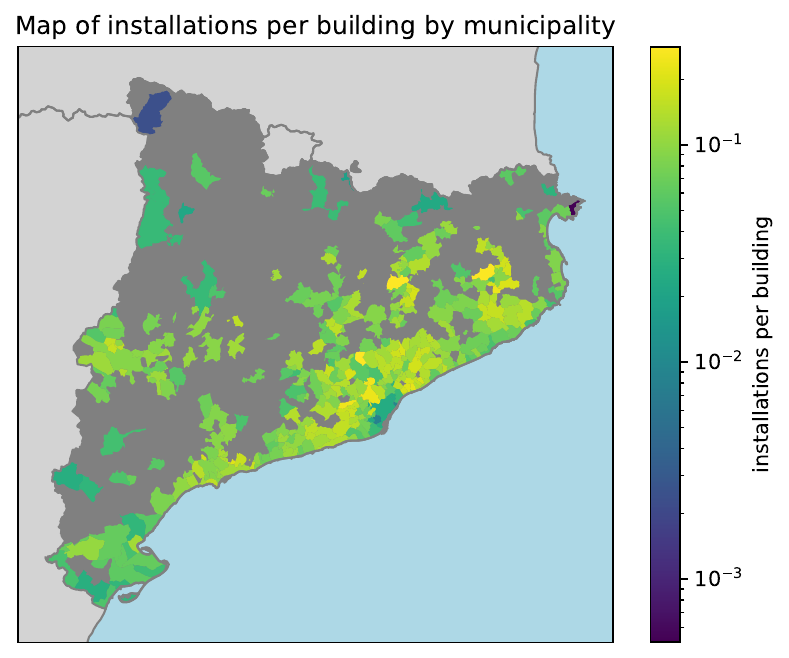} 
\end{center}
\caption{Installations per building by municipality with a logarithmic color scale. The municipalities with no building data or no installations are displayed in dark gray.}
\label{sfigure6}
\end{figure}

While in single-family housing one would expect the number of installations to scale approximately linearly with the population, this relationship may differ in apartment buildings, which contain multiple dwellings. In such cases, installations are typically undertaken at the building level rather than by individual households, so the number of installations may scale more closely with the number of buildings than with the population. For this reason, installations per building can be a more appropriate metric than installations per capita. However, this approach requires reliable data on the number of buildings, which is not readily available for smaller municipalities. The results for the municipalities where such data is available are shown in Fig.~\ref{sfigure6}. They are qualitatively consistent with the spatial distribution obtained using installations per capita, and the main conclusions remain unchanged.

\section{Spatial heterogeneity in \textit{el Barcelonès}}
\label{supplementary:bcn_neighbourhoods}

To analyze in greater detail the relationship between installations per capita and other variables, we focus on the municipality of Barcelona. In this case, the analysis can be refined by using neighborhoods as the spatial unit (73 in total, maps from \cite{mapes_barris}), allowing us to exploit variations in population density and income to assess whether the patterns observed at the municipal level also hold at smaller scales. 

Overall, installations per capita appear to be significantly associated with both housing type and socioeconomic status. However, establishing a direct causal relationship with either factor is challenging, as the observed correlations may be mediated by other variables. In addition, the dataset is predominantly urban, which likely limits the representation of behaviors associated with single-family housing. 

\begin{figure}[h]
\begin{center}
    \includegraphics[width=\columnwidth]{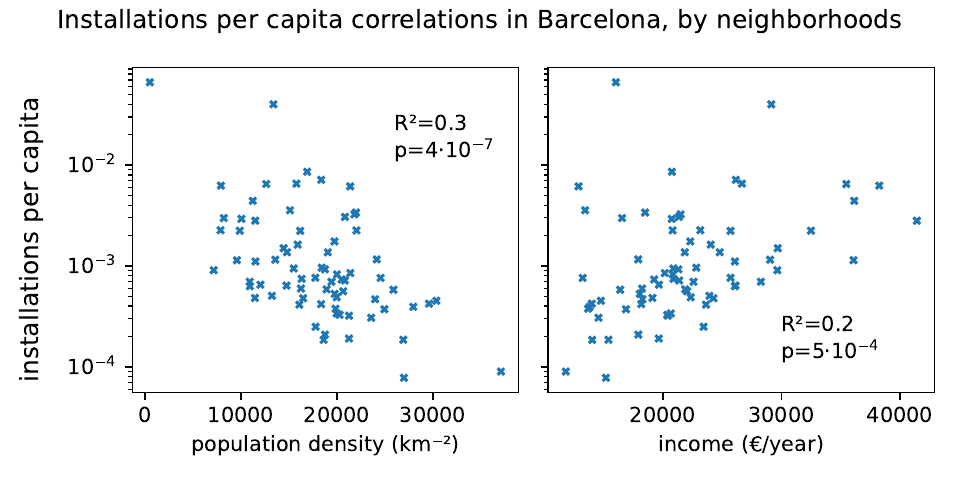}
\end{center}
\caption{Scatter plot of the installations per capita and the population density (considering only above ground constructed surface), at the left side, and the household disposable income, at the right side, both in Barcelona, by neighborhoods.}
\label{sfigure7}
\end{figure}

To further investigate these relationships, we analyze the correlation between the logarithm of installations per capita and both population density and household disposable income \cite{bcnrdlpc}, as shown in Fig.~\ref{sfigure7}. For population density, we find a negative correlation with a coefficient of $R^{2} = 0.3$ and a statistically significant p-value. In contrast, income exhibits a positive correlation, with $R^{2} = 0.2$, also statistically significant.

These results are particularly informative due to the finer spatial resolution compared to the municipality-level analysis, where the corresponding coefficients were $R^{2} = 0.2$ for urban population density and $R^{2} = 0.08$ for income. The higher resolution reveals stronger correlations, likely because local variations are not averaged out as they are at larger spatial scales.

Finally, the stronger correlation with population density than with income suggests that housing characteristics may play a more dominant role than economic factors in shaping PV adoption, in line with the findings presented in the main article.

\end{document}